\documentclass[final,3p,times]{elsarticle}

\usepackage{amssymb}

\usepackage{graphicx}
\usepackage{subcaption}
\usepackage{caption}
\graphicspath{ {images/} }

\usepackage{dcolumn}

\usepackage{bm}

\usepackage[normalem]{ulem}
\useunder{\uline}{\ul}{}

\usepackage{lipsum}
\usepackage{natbib}
\usepackage{url}

\usepackage{multirow}

\biboptions{square,numbers,sort&compress}

\begin{document}

\begin{frontmatter}



\title{Analysis on the urban street network of Korea: Connections between topology and meta-information}
\author{Byoung-Hwa Lee}
\address{Department of Physics, Pohang University of Science and Technology, Pohang 37673, Republic of Korea}
\author{Woo-Sung Jung*}
\address{Department of Physics, Pohang University of Science and Technology, Pohang 37673, Republic of Korea}
\address{Department of Industrial and Management Engineering, Pohang University of Science and Technology, Pohang 37673, Republic of Korea}
\address{Asia Pacific Center for Theoretical Physics, Pohang 37673, Republic of Korea}


\begin{abstract}
Cities consist of infrastructure that enables transportation, which can be considered as topology in abstract terms. Once cities are physically organized in terms of infrastructure, people interact with each other to form the values, which can be regarded as the meta-information of the cities. The topology and meta-information coevolve together as the cities are developed. In this study, we investigate the relationship between the topology and meta-information for a street network, which has aspects of both a complex network and planar graph. The degree of organization of a street structure determines the efficiency and productivity of the city in that they act as blood vessels to transport people, goods, and information. We analyze the topological aspect of a street network using centralities including the betweenness, closeness, straightness, and information. We classify the cities into several groups that share common meta-information based on the centrality, indicating that the topological factor of the street structure is closely related to meta-information through coevolution. We also obtain the coevolution in the planned cities using the regularity. Another footprint is the relation between the street segment length and the population, which shows the sublinear scaling.
\end{abstract}


\end{frontmatter}



\section{Introduction}
Cities have become a very important part of our lives with the global trends towards rapid urbanization. As urban areas have increased in size the transportation of people, goods, and information within or between cities has been extensively studied in diverse disciplines, from transportation engineering to complex network theory. Of the many means of transportation within cities, automobiles are the most commonly used vehicles compared to others such as trains, airplanes, or ships. The street network, which is a kind of blood vessel for cars, constitutes the backbone of the cities.

Over the last two decades, the use of complex network theory has been successful in understanding social, economic, and biological issues \cite{newman2003structure, albert2002statistical, castellano2009statistical}. They also offer a powerful methodology for the analysis of street networks. In this sense, many studies have analyzed street networks from the perspective of complex networks \cite{barthelemy2011spatial}. One main approach in these studies is the centrality analysis using the empirical approach \cite{crucitti2006centrality1, porta2006network1, porta2006network2, buhl2006topological, lammer2006scaling}. Two distinctive issues of street network analysis are the topological perspective with centrality and network efficiency defined using the shortest path \cite{latora2001efficient}. Cities are classified using centrality measures \cite{crucitti2006centrality2} or network efficiency \cite{cardillo2006structural}. Another focus of the study is the modeling of a street network \cite{barthelemy2008modeling, courtat2011mathematics, perna2011characterization}. Using basic rules that pertain to the intrinsic organization mechanism of streets, street patterns can be modeled successfully, and their topological properties are the same as that of a real street structure. The organization mechanism is also related to the temporal evolution of the street network \cite{strano2012elementary, barthelemy2013self}.

In this study, we analyzed the urban street network of 22 cities in Korea. Each city has its own distinctive properties, and hence its street structures are diverse. For example, cities have different roles, and there include capital, satellite, farming, or industrial specialized cities. An industrial specialized city has a different street structure compared to a farming town. In addition, cities have different histories. Some cities were founded in the medieval era, while others were established in the modern era. Medieval cities have irregular street patterns with short or crooked streets, while modern cities have a hierarchical structure with regular, rectilinear, orthogonal, or tree-like patterns \cite{marshall2004streets}. The topography is another factor that characterizes the street pattern; cities in flatlands, mountains, or by the seaside have their own distinctive street structures, which are determined by geographical constraints. Different cultural, social, and economic backgrounds determine the street patterns of a given city. For instance, most Americans prefer detached houses as opposed to apartments, but most Koreans have the opposite taste.

The meta-information is basically information about information. Although the city has the physical structure, it is also composed of various kinds of information. For example, a person is the element of the city and is also information in the document. However, a whole collection of people, i.e. the population, is the meta-information because it is information about information. Above mentioned four properties are also kinds of the meta-information. Other types of meta-information include the area of the city, the gross regional domestic product (GRDP), the GRDP per capita, the total or average street segment length, the regularity, and whether or not have special kinds of administrative properties such as city planning or the urban-rural integrated city which will be explained in the last section. Unlike the meta-information, the topology is determined by only the basic network quantities, such as the degree or centrality. The external factors of the cities, such as history, topography, or population, could simultaneously function as both the cause and effect for the topology of the street network. Then, the meta-information and topology are coevolved throughout the organization process of the cities. As a result, the classification based on the topology would be related to the meta-information. In this study, we conducted an investigation to determine their relationship using the centrality, regularity, and scaling relation.

This paper is structured as follows. First, we show the results of the centrality analysis and classification of the cities. In this first section, we present the relationship between the topology and meta-information directly. Then, in the subsequent two sections, regularity and the scaling law in the street system reveal the implicit connection between the topology and meta-information. Finally, we summarize the results and discuss future studies.

\section{Data}

\begin{table*}[!t]
\centering
\footnotesize
\caption{Basic information of 22 Korean cities. $N$ is the number of nodes and $L$ is the number of links. $\langle k \rangle$ is the average degree and $\sigma_{k}$ is the standard deviation of the degrees. The Tot. Length is the total street segment length in $\mathrm{km}$. The Avg. Length is the average street segment length in $\mathrm{m}$. An area is in $\mathrm{{km}^2}$. GRDP is the gross regional domestic product in trillions of the Korean won. GRDP per capita is in millions of the Korean won. The Urb.-Rur. Int. is the urban-rural integrated city explained in the text.}
\label{table:stat}
\begin{tabular}{crrrrrc}
\hline
\multirow{2}{*}{}            & \multicolumn{1}{c}{N} & \multicolumn{1}{c}{$\langle k \rangle$} & \multicolumn{1}{c}{Tot. Length} & \multicolumn{1}{c}{Population} & \multicolumn{1}{c}{GRDP}            & Metropolitan           \\
                             & \multicolumn{1}{c}{L} & \multicolumn{1}{c}{$\sigma_{k}$}        & \multicolumn{1}{c}{Avg. Length} & \multicolumn{1}{c}{Area}       & \multicolumn{1}{c}{per capita} & Urb.-Rur. Int. \\ \hline
\multirow{2}{*}{Seoul}       & 8,582                 & 3.37                                    & 3372.7                          & 10,103,230                     & 318.6                               & Yes                    \\
                             & 14,471                & 1.08                                    & 233.1                           & 605.2                          & 32.1                                & No                     \\ \hline
\multirow{2}{*}{Busan}       & 2,563                 & 3.02                                    & 1430.3                          & 3,557,716                      & 70.3                                & Yes                    \\
                             & 3,865                 & 0.85                                    & 370.1                           & 765.9                          & 20.5                                & No                     \\ \hline
\multirow{2}{*}{Incheon}     & 2,062                 & 2.97                                    & 1212.8                          & 2,985,831                      & 64.7                                & Yes                    \\
                             & 3,063                 & 0.84                                    & 395.9                           & 1040.9                         & 22.9                                & No                     \\ \hline
\multirow{2}{*}{Daegu}       & 1,419                 & 3.16                                    & 1072.5                          & 2,518,467                      & 44.8                                & Yes                    \\
                             & 2,243                 & 0.80                                    & 478.2                           & 883.6                          & 18.2                                & No                     \\ \hline
\multirow{2}{*}{Daejeon}     & 2,153                 & 2.90                                    & 1055.8                          & 1,551,931                      & 31.5                                & Yes                    \\
                             & 3,118                 & 0.85                                    & 338.6                           & 540.1                          & 20.4                                & No                     \\ \hline
\multirow{2}{*}{Gwangju}     & 2,620                 & 2.97                                    & 1127.5                          & 1,492,948                      & 29.8                                & Yes                    \\
                             & 3,886                 & 0.85                                    & 290.2                           & 501.2                          & 19.7                                & No                     \\ \hline
\multirow{2}{*}{Ulsan}       & 2,636                 & 2.87                                    & 1224.8                          & 1,192,262                      & 68.3                                & Yes                    \\
                             & 3,780                 & 0.85                                    & 324.0                           & 1060.2                         & 60.6                                & No                     \\ \hline
\multirow{2}{*}{Suwon}       & 955                   & 2.90                                    & 405.6                           & 1,213,665                      & 25.0                                & No                     \\
                             & 1,385                 & 0.88                                    & 292.9                           & 121.0                          & 22.1                                & No                     \\ \hline
\multirow{2}{*}{Goyang}      & 1,073                 & 2.81                                    & 493.6                           & 1,028,237                      & 15.6                                & No                     \\
                             & 1,505                 & 0.87                                    & 328.0                           & 267.4                          & 15.9                                & No                     \\ \hline
\multirow{2}{*}{Bucheon}     & 491                   & 2.87                                    & 189.4                           & 849,064                        & 15.6                                & No                     \\
                             & 704                   & 0.83                                    & 269.1                           & 53.5                           & 18.1                                & No                     \\ \hline
\multirow{2}{*}{Ansan}       & 1,013                 & 3.16                                    & 459.6                           & 696,934                        & 22.6                                & No                     \\
                             & 1,601                 & 0.78                                    & 287.0                           & 149.5                          & 31.7                                & No                     \\ \hline
\multirow{2}{*}{Jeonju}      & 1,064                 & 2.98                                    & 462.4                           & 652,858                        & 10.2                                & No                     \\
                             & 1,585                 & 0.91                                    & 291.7                           & 206.1                          & 15.8                                & No                     \\ \hline
\multirow{2}{*}{Pohang}      & 1,831                 & 2.74                                    & 845.6                           & 519,244                        & 17.9                                & No                     \\
                             & 2,505                 & 0.93                                    & 337.6                           & 1128.8                         & 35.2                                & Yes                    \\ \hline
\multirow{2}{*}{Gwangmyeong} & 206                   & 2.57                                    & 92.0                            & 346,888                        & 5.8                                 & No                     \\
                             & 265                   & 0.93                                    & 347.3                           & 38.5                           & 16.5                                & No                     \\ \hline
\multirow{2}{*}{Chuncheon}   & 809                   & 2.47                                    & 538.9                           & 281,005                        & 5.4                                 & No                     \\
                             & 1,000                 & 0.85                                    & 538.9                           & 1116.4                         & 19.9                                & Yes                    \\ \hline
\multirow{2}{*}{Gyeongju}    & 942                   & 2.74                                    & 777.2                           & 261,535                        & 8.3                                 & No                     \\
                             & 1,291                 & 0.88                                    & 602.1                           & 1324.4                         & 32.9                                & Yes                    \\ \hline
\multirow{2}{*}{Guri}        & 212                   & 2.71                                    & 79.4                            & 186,774                        & 3.4                                 & No                     \\
                             & 287                   & 0.93                                    & 276.6                           & 33.3                           & 17.6                                & No                     \\ \hline
\multirow{2}{*}{Gimcheon}    & 442                   & 2.53                                    & 409.2                           & 140,085                        & 3.3                                 & No                     \\
                             & 559                   & 0.78                                    & 732.0                           & 1009.5                         & 26.0                                & Yes                    \\ \hline
\multirow{2}{*}{Gongju}      & 496                   & 2.21                                    & 500.8                           & 113,294                        & 2.9                                 & No                     \\
                             & 549                   & 0.71                                    & 912.3                           & 864.3                          & 23.7                                & Yes                    \\ \hline
\multirow{2}{*}{Donghae}     & 74                    & 2.05                                    & 49.8                            & 94,562                         & 2.4                                 & No                     \\
                             & 76                    & 0.49                                    & 655.7                           & 180.0                          & 26.2                                & No                     \\ \hline
\multirow{2}{*}{Namwon}      & 314                   & 2.17                                    & 323.4                           & 84,856                         & 1.5                                 & No                     \\
                             & 341                   & 0.74                                    & 948.4                           & 752.6                          & 19.5                                & Yes                    \\ \hline
\multirow{2}{*}{Gwacheon}    & 105                   & 2.80                                    & 55.2                            & 69,031                         & 2.6                                 & No                     \\
                             & 147                   & 0.93                                    & 375.5                           & 35.9                           & 37.3                                & No                     \\ \hline
\multirow{2}{*}{Average}     & 1457                  & 2.77                                    & 735.4                           & 1,360,928                      & 35.0                                & \multirow{2}{*}{}      \\
                             & 2192                  & 0.84                                    & 437.5                           & 576.3                          & 25.1                                &                        \\ \hline
\end{tabular}
\end{table*}

We employed the Standard Node-Link database provided by the Ministry of Land, Infrastructure, and Transport of Korea \cite{standardnodelink}. In this database, identification numbers are assigned to almost all of the roads and intersections in Korea, ranging from small roads to highways. The criterion for choosing a road is that the roads should have over two lanes with a center line; thus, very small roads such as alleys or driveways, which do not significantly affect to the whole traffic flow, are excluded in this database. Intersections and roads correspond to nodes and links, respectively, in terms of the primal network. Although both ends of a bridge, tunnel, overpass, or underpass are also regarded as an intersection in the database, the proportion of these types of nodes is negligible compared to the nodes for the intersections. The total number of nodes and links in Korea is about 100,000 and 130,000, respectively. The Standard Node-Link database also contains the geographical coordinates of the nodes, as well as the actual lengths of the roads.

In Korea, there are seven metropolitan cities and 78 other cities for the administrative district. We analyzed the street networks of the seven metropolitan cities and 15 non-metropolitan cities, so for a total of 22 cities.  The criterion for selection of the non-metropolitan cities is basically random samples from a table of the whole cities in descending order of the population. The cities that were analyzed have diverse sizes, roles, topographies, and characteristics; the populations range from 70k to 10M, and the areas range from 33 $\mathrm{{km}^2}$ to 1,324 $\mathrm{{km}^2}$, which is an order of magnitude of about 40 times when comparing the minimum with the maximum. Of the 15 non-metropolitan cities, seven cities are located in the capital region, six cities are adjacent to the sea, and six cities are urban-rural integrated cities. Some of the cities have mountainous area, and other cities are located in flatlands or basins. One of the cities, Incheon, even contains several islands. The diversity of the characteristics in the cities is related to the street structure, and we investigated the street network using methods for complex networks.

\section{Centrality Analysis}
Below, we analyze the structure of a street network using the centrality, which quantifies the importance of certain nodes or links within the network, and which is widely adopted as an important method in the field of social sciences, data sciences, and physics. We employed the four kinds of centrality: betweenness, closeness, straightness, and information centrality.

The betweenness centrality is one of the primary centrality in social networks \cite{freeman1978centrality}, and indicates the significance of a node or link in terms of the shortest paths. The betweenness centrality of a node $i$ in an undirected weighted graph $G$ is defined as
\begin{equation}
C_{B,i}=\frac{1}{(N-1)(N-2)}\sum_{j, k \in G, j \neq k \neq i} \frac{\sigma_{jk} (i)}{\sigma_{jk}},
\end{equation}
where $\sigma_{jk}$ is the number of shortest paths between $j$ and $k$, and $\sigma_{jk}(i)$ is the number of shortest paths between $j$ and $k$ that pass node $i$. The term, $1/(N-1)(N-2)$, refers to the normalization such that $C_{B,i}\in[0,1]$, where $N$ is the number of nodes in the network. The shortest paths are determined by minimization of the sum of the weights of links, which is the Euclidean distance.

In the case where all of the vehicles on a street network follow the rule that they have to pass by the node or link only along with the shortest path, then the betweenness centrality approximately represents the volume of traffic. Another required assumption is that all of the nodes and links have an equivalent volume of vehicles departing from that node or link.

The closeness centrality measures the degree of proximity to all other nodes in a network, and is defined as \cite{sabidussi1966centrality, wasserman1994social}
\begin{equation}
C_{C,i}=\frac{N-1}{\sum\limits_{j \in G, j \neq i} d_{ij}},
\end{equation}
where $d_{ij}$ is the shortest path length between node $i$ and $j$. The closeness centrality is also calculated from the Euclidean distance, and not the chemical distance. 

The straightness centrality quantifies the deviation from a virtual straight route, and is defined as \cite{latora2001efficient}
\begin{equation}
C_{S,i}=\frac{1}{N-1}\sum_{j \in G, j \neq i} \frac{d_{ij}^{\mathrm{straight}}}{d_{ij}},
\end{equation}
where $d_{ij}^{\mathrm{straight}}$ is the straight-line Euclidean distance between node $i$ and $j$, and $d_{ij}$ is the actual Euclidean distance between nodes $i$ and $j$. For the normalization, we divided by $(N-1)$. If one route is aligned well with the other routes, then the corresponding node would have a high value of straightness centrality.

The information centrality \cite{latora2004measure} measures the decreasing amount of the efficiency when the specific node is deactivated. In the deactivation process, the node remains but the links connected to that node are removed. The information centrality of node $i$ is defined by the relative decrement of the network efficiency by the deactivation from $G$ of the links incident in node $i$,
\begin{equation}
C_{I,i}=\frac{\Delta E}{E}=\frac{E[G]-E[G']}{E[G]},
\end{equation}
where $E[G]$ is the efficiency of an original graph $G$ and $E[G']$ is the decreased efficiency of the graph $G'$ that the node $i$ is removed. The efficiency of a graph $G$ is defined as \cite{latora2001efficient, latora2005vulnerability}
\begin{equation}
E[G]=\frac{1}{N(N-1)}\sum_{i,j \in G, j \neq i} \frac{d_{ij}^{\mathrm{straight}}}{d_{ij}},
\end{equation}
where $d_{ij}^{\mathrm{straight}}$ is the straight-line Euclidean distance between node $i$ and $j$ and $G'$ is the graph with $N$ nodes and $K-k_{i}$ links by deleting the links incident in node $i$ from the original graph $G$.

\begin{figure}[!t]
\includegraphics[width=1\textwidth]{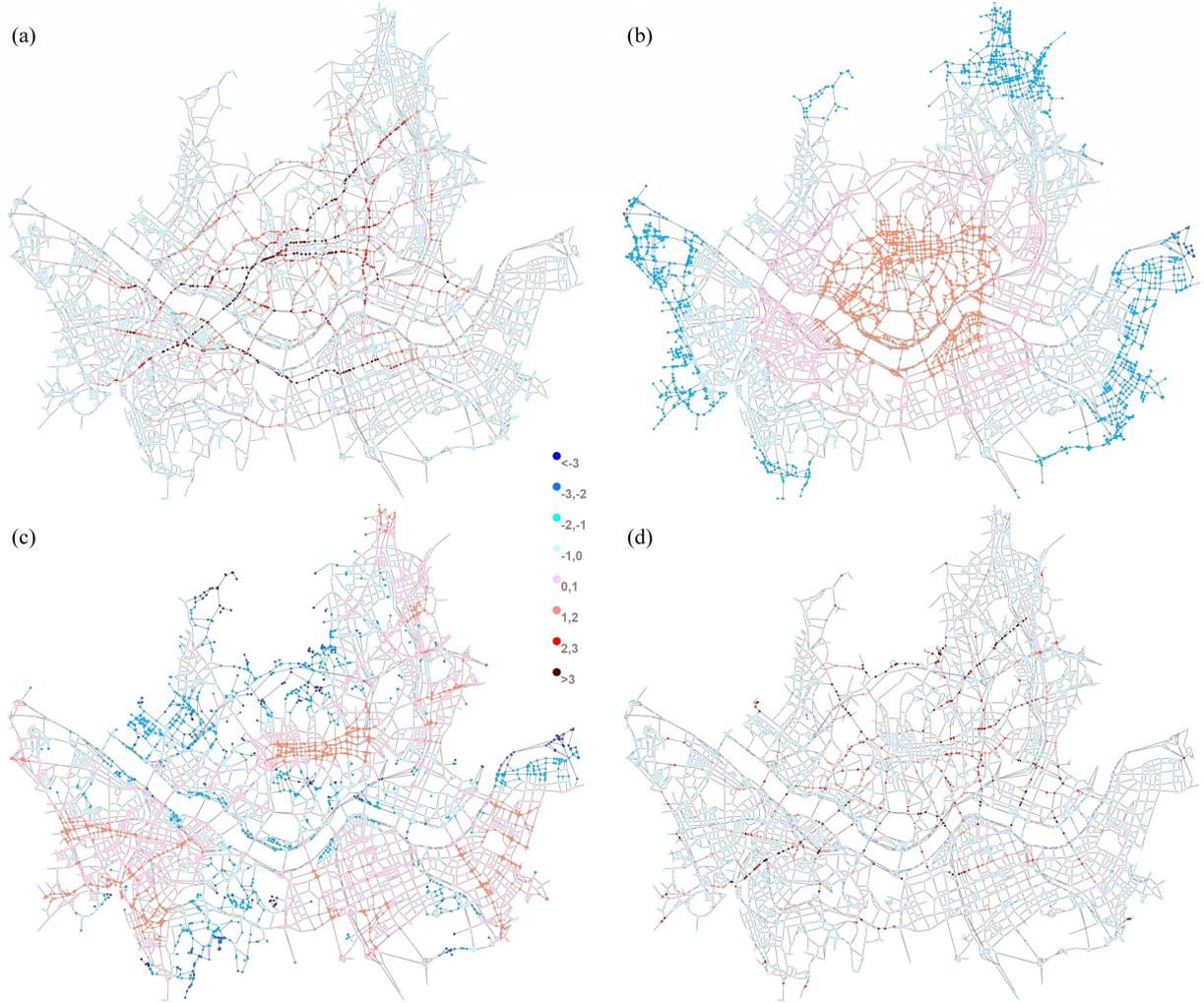}\centering
\caption{(Color online) The centrality maps for Seoul. (a) The betweenness centrality map shows high heterogeneity since some nodes have extremely large values. (b) The closeness centrality map shows that the nodes located in the geographical center of the city tend to have the large value of the closeness centrality, but its discrepancy is small, which means homogeneity. (c) The straightness centrality map shows that the nodes along with the main axes have high values, but has homogeneity compared to the betweenness and information centrality. (d) The information centrality map shows that some nodes have extremely high values and has medium heterogeneity.}
\label{fig:seoul}
\end{figure}

After calculating the centralities of each node, we plotted the maps of centralities for each city. Fig. \ref{fig:seoul} represents the example of Seoul for four centralities. According to the values deviated from the average of each city, the nodes have different colors. For example, if the centrality value of a node is $0.5\sigma$ larger than the average, then the color of the node is light magenta. The more central nodes tend to have large closeness centrality, and the less angularly deviated nodes from the primary axes tend to have large straightness centrality. The closeness and straightness centrality are less spatially heterogeneous, while the betweenness and information centrality are more spatially heterogeneous because they have extreme values larger than $3\sigma$.

The extremely high values of the betweenness centrality are related to the large traffic volumes, which means that they are vulnerable to traffic congestion. For instance, the streets in the central business district of Seoul have a high betweenness centrality, which is consistent with the fact that the traffic volume of that area is particularly high compared to the other regions in Seoul. The connection between the betweenness centrality and the traffic volume is also an interesting topic, but we leave it for the further study.

\begin{figure*}[!t]\centering
\begin{subfigure}[t]{0.3\textwidth}\centering\caption{}
\includegraphics[width=\textwidth]{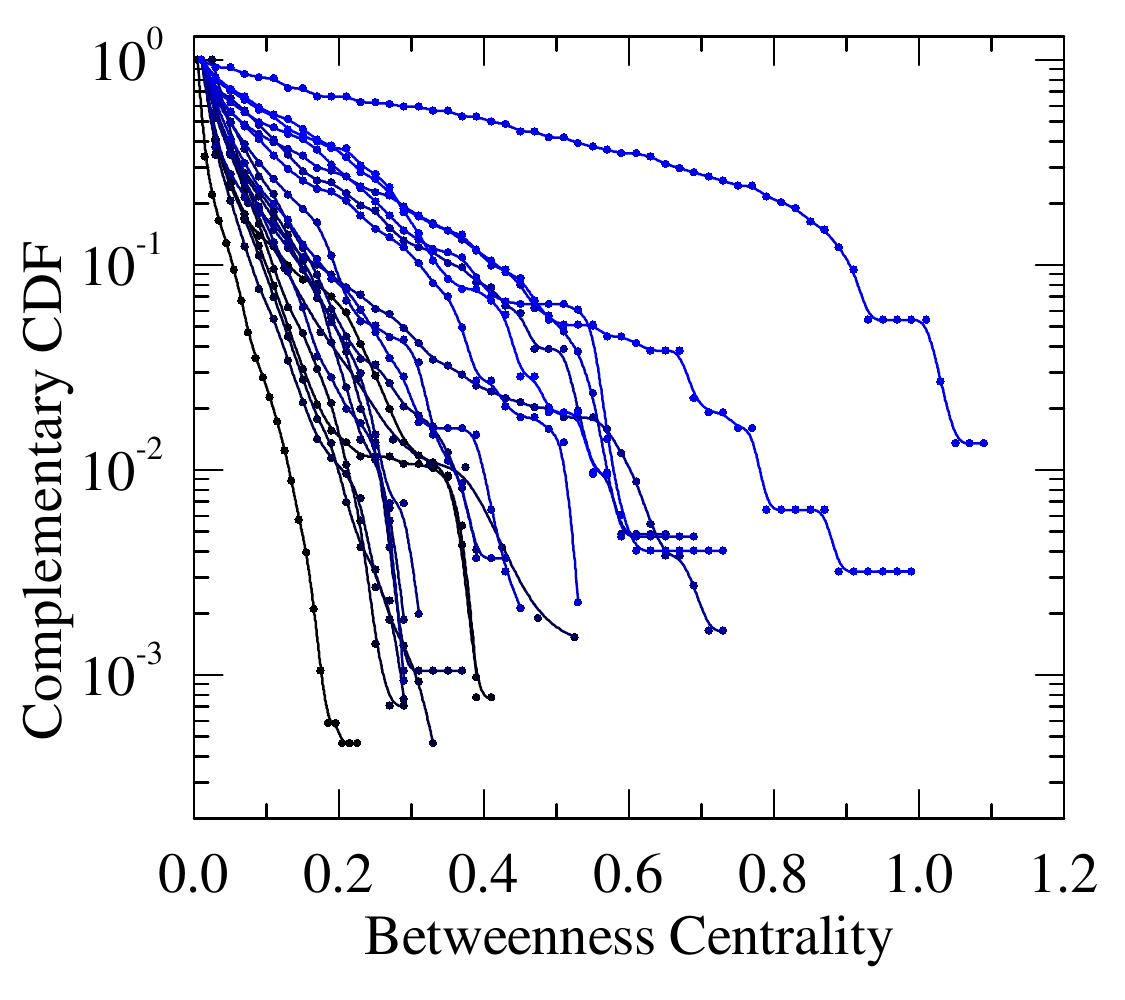}\label{ccdfbc}
\end{subfigure}\quad
\begin{subfigure}[t]{0.3\textwidth}\centering\caption{}
\includegraphics[width=\textwidth]{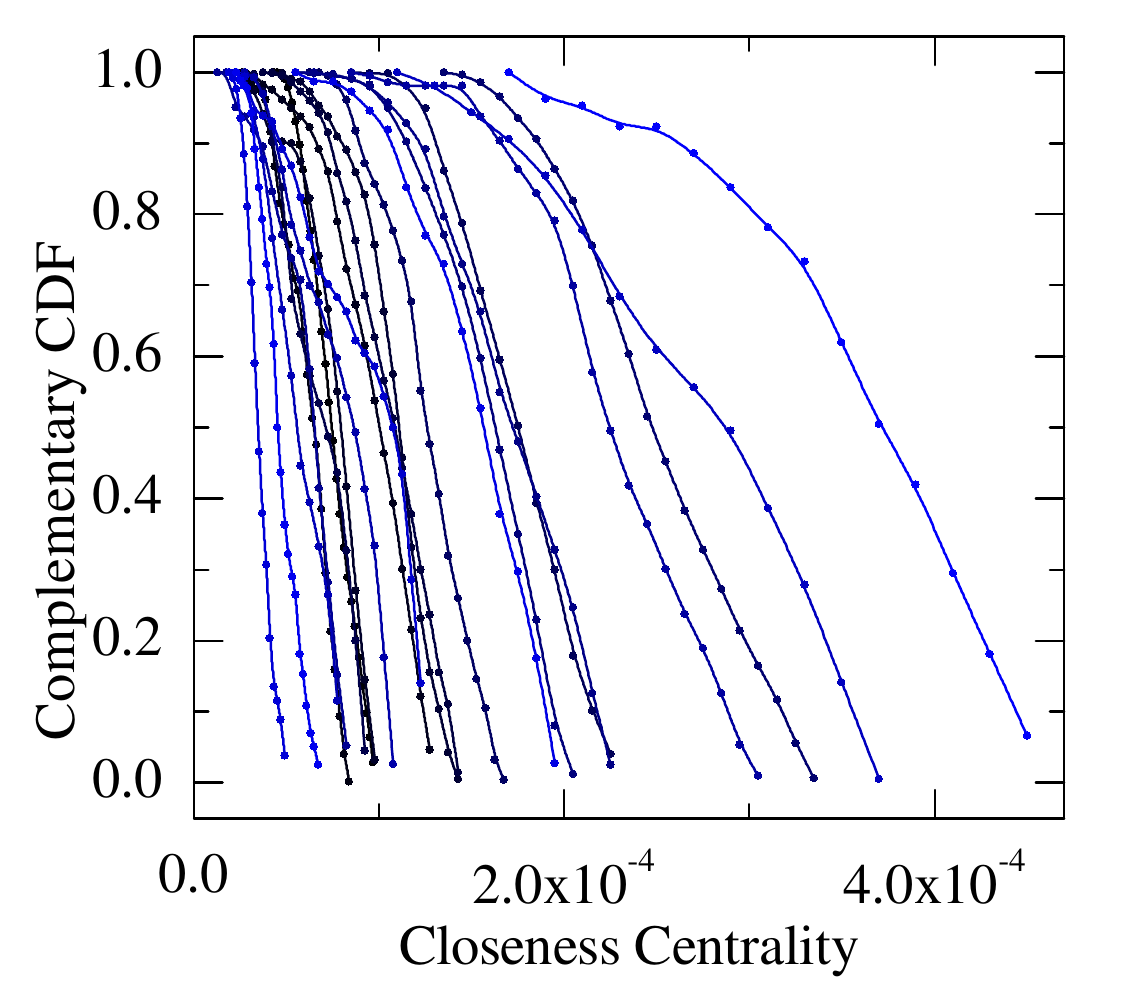}\label{ccdfcc}
\end{subfigure}\\
\begin{subfigure}[t]{0.3\textwidth}\centering\caption{}
\includegraphics[width=\textwidth]{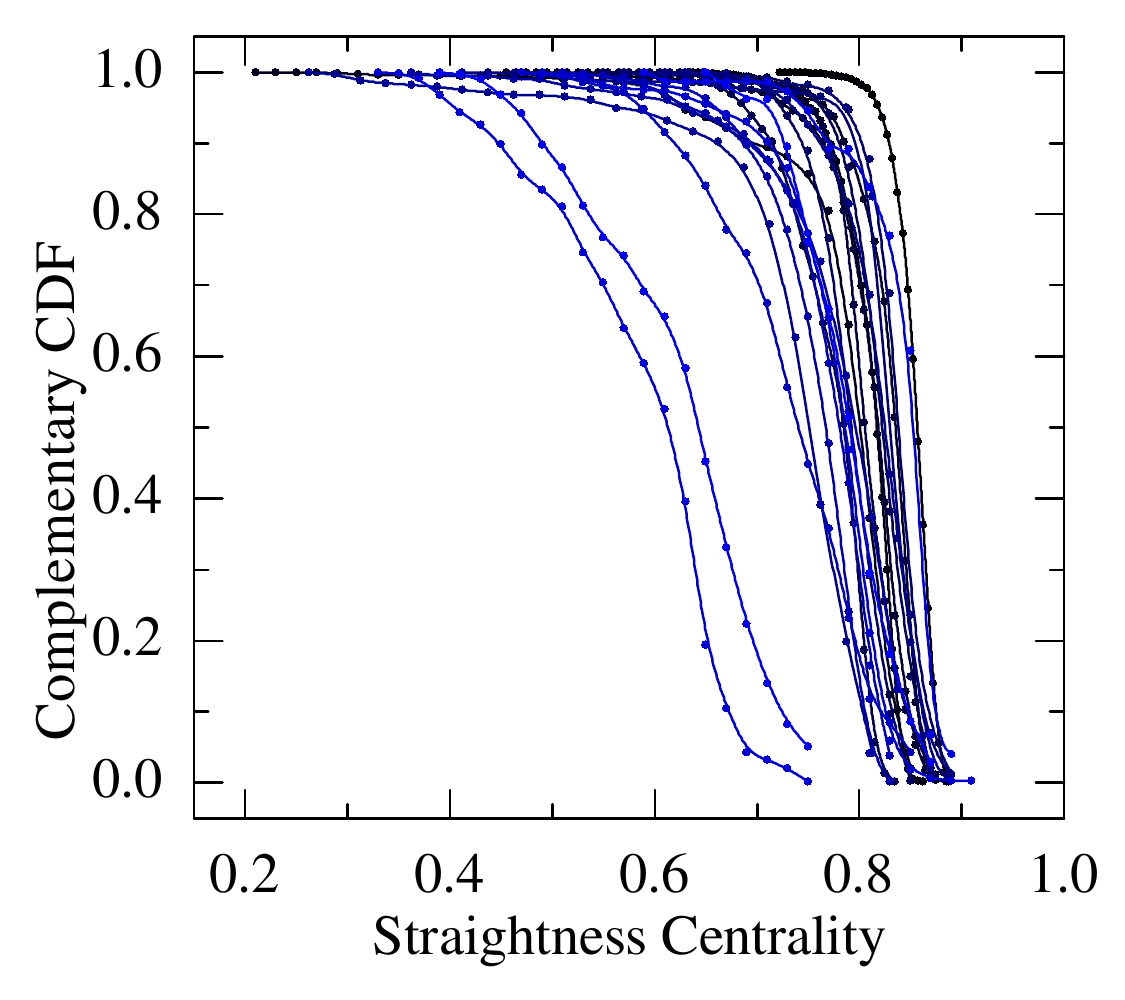}\label{ccdfsc}
\end{subfigure}\quad
\begin{subfigure}[t]{0.3\textwidth}\centering\caption{}
\includegraphics[width=\textwidth]{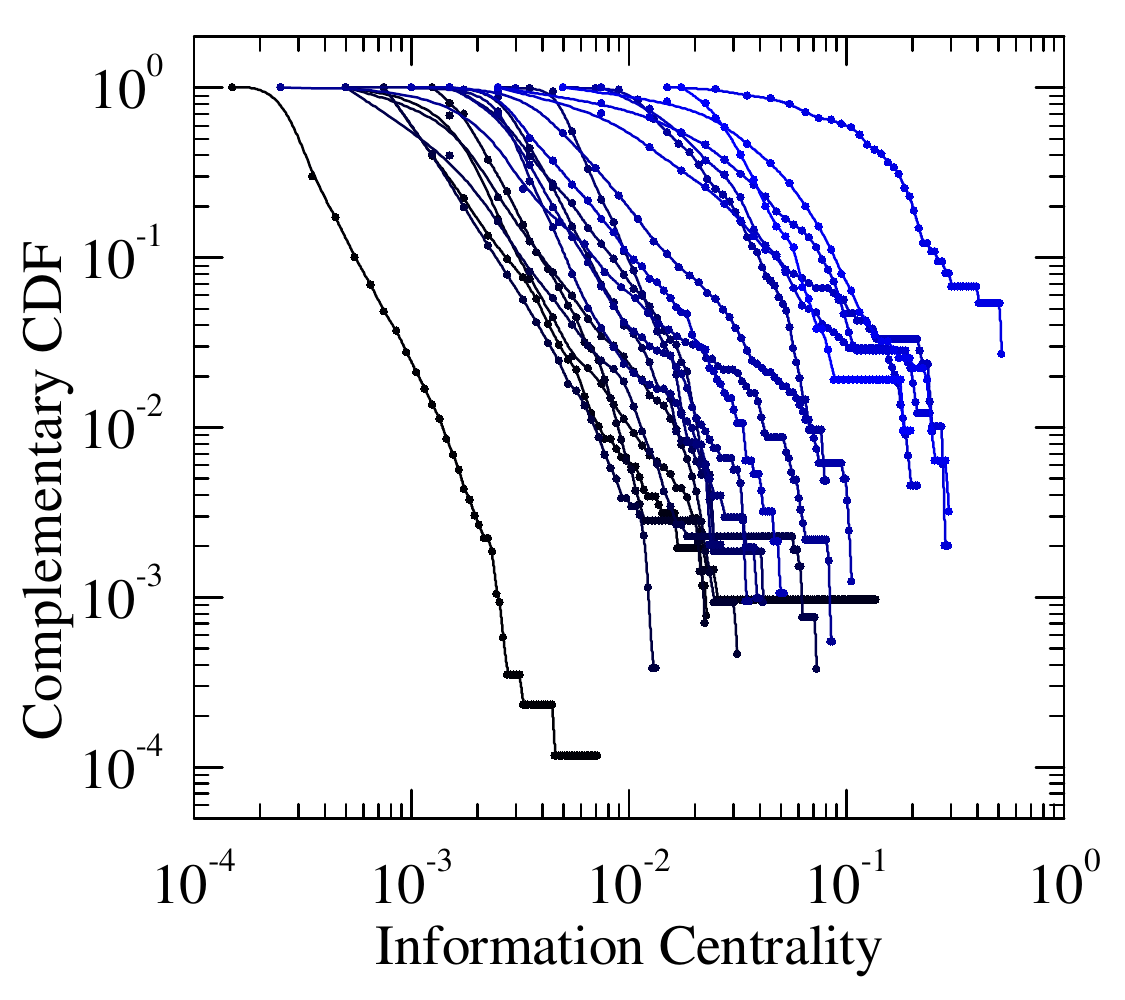}\label{ccdfic}
\end{subfigure}
\caption{(Color online) The complementary cumulative distribution function (CCDF) of four centralities. (a) The CCDF of the betweenness centrality has the exponential distribution with the diverse slopes between the cities. (b) The CCDF of the closeness centrality is linearly decreasing function except the head, which means that the probability distribution function is evenly distributed. (c) The probability distribution function of the straightness centrality is an unimodal distribution, indicating that the value is concentrated in the center (0.8). (d) The CCDF of the information centrality is partially power-law.}
\label{fig:dist}
\end{figure*}

We plotted the complementary cumulative distribution function of each centrality. These results are illustrated in Fig. \ref{fig:dist} (a - d). The betweenness centrality has an exponential distribution with some noises in the tail, and their exponents are different from each other. The discrepancy between the slopes may stem from the fact that the cities we selected have a large range of sizes, from $N=74$ to $N=8582$, despite the normalization in Eq. (1). The probability distribution function for the closeness centrality is almost evenly distributed, with the exception of the head. The probability distribution function of the straightness centrality is unimodal centered approximately at 0.8, which means that most of its value is similar to each other. The CCDF of the information centrality is partially power-law. For large cities such as the metropolitan cities, the CCDF is the power-law in the middle range, however, for the small cities, the CCDF is more like the exponential distribution. Of the four kinds of centrality, the betweenness and information centrality are closer to the concept of the centrality, since the centrality is generally more meaningful for a great diversity in their values.

Heterogeneity in the betweenness centrality distribution corresponds to the fact that the city could have a traffic problem compared to the case where it has a homogeneous betweenness centrality distribution. Here, we employed the Gini coefficient to diagnose the traffic problem in terms of topology.

\section{Classification of the Cities}
The Gini coefficient, widely used in economics, measures the inequality of economic quantities such as the income. If $x_{i}$ is the value of the agent $i$, and there are $n$ agents, then the Gini coefficient, $g$, is defined as
\begin{equation}
g=\frac{\sum_{i=1}^{n}\sum_{j=1}^{n}|x_{i}-x_{j}|}{2n\sum_{i=1}^{n}x_{i}}.
\end{equation}
The two extreme values, 0 and 1, correspond to the perfect equality and maximal inequality, respectively. Here, we consider $x_{i}$ as the centrality value of node $i$. We calculate its Gini coefficient to represent the heterogeneity in only one value. The Gini coefficients of the cities are shown in Table \ref{table:gini}. The heterogeneity of the betweenness centrality confirmed in the probability distribution function in Fig. \ref{fig:dist} (a) is also valid for the Gini coefficient analysis. The Gini coefficient for the betweenness centrality is larger than that for the other centralities, which means the huge inequality. The information centrality has medium Gini coefficient. The closeness and straightness centrality have smaller Gini coefficients, and their smaller inequality originated from the planarity of the street network.

\begin{table*}[!t]
\centering
\small
\caption{The Gini coefficients of four kinds of centrality: betweenness ($g_{BC}$), closeness ($g_{CC}$), straightness ($g_{SC}$), and information ($g_{IC}$).}
\label{table:gini}
\begin{tabular}{crrrr}
\hline
            & \multicolumn{1}{c}{$g_{BC}$} & \multicolumn{1}{c}{$g_{CC}$} & \multicolumn{1}{c}{$g_{SC}$} & \multicolumn{1}{c}{$g_{IC}$} \\ \hline
Seoul       & 0.71                         & 0.10                         & 0.012                        & 0.23                         \\
Busan       & 0.73                         & 0.13                         & 0.019                        & 0.35                         \\
Incheon     & 0.66                         & 0.13                         & 0.038                        & 0.37                         \\
Daegu       & 0.58                         & 0.14                         & 0.020                        & 0.27                         \\
Daejeon     & 0.63                         & 0.13                         & 0.026                        & 0.42                         \\
Gwangju     & 0.65                         & 0.10                         & 0.022                        & 0.31                         \\
Ulsan       & 0.72                         & 0.16                         & 0.028                        & 0.44                         \\
Suwon       & 0.61                         & 0.11                         & 0.020                        & 0.30                         \\
Goyang      & 0.61                         & 0.11                         & 0.024                        & 0.39                         \\
Bucheon     & 0.55                         & 0.11                         & 0.018                        & 0.21                         \\
Ansan       & 0.60                         & 0.11                         & 0.022                        & 0.28                         \\
Jeonju      & 0.60                         & 0.13                         & 0.026                        & 0.35                         \\
Pohang      & 0.76                         & 0.16                         & 0.054                        & 0.58                         \\
Gwangmyeong & 0.58                         & 0.12                         & 0.046                        & 0.33                         \\
Chuncheon   & 0.59                         & 0.17                         & 0.036                        & 0.50                         \\
Gyeongju    & 0.62                         & 0.17                         & 0.036                        & 0.43                         \\
Guri        & 0.62                         & 0.15                         & 0.059                        & 0.51                         \\
Gimcheon    & 0.60                         & 0.19                         & 0.037                        & 0.58                         \\
Gongju      & 0.58                         & 0.12                         & 0.083                        & 0.56                         \\
Donghae     & 0.43                         & 0.12                         & 0.025                        & 0.41                         \\
Namwon      & 0.53                         & 0.13                         & 0.078                        & 0.47                         \\
Gwacheon    & 0.50                         & 0.11                         & 0.033                        & 0.29                         \\ \hline
Average     & 0.61                         & 0.13                         & 0.035                        & 0.39                         \\ \hline
\end{tabular}
\end{table*}

After obtaining the Gini coefficients for four kinds of centrality, entire cities are classified topologically \cite{crucitti2006centrality1}. The four Gini coefficients of the centrality for each city represent the axes in the four-dimensional space made by the centralities. Each city corresponds to one point in that space. Then, the Euclidean distance between cities in the centrality space indicates how similar they are to each other in terms of the network topology. In this case, the distance is not exactly the actual Euclidean distance; instead, the distance between two cities $m$ and $n$, $D_{mn}$, is defined as
\begin{equation}
D_{mn}=\Bigg[\sum_{i=1}^{4}\frac{(g^{m}_{i} - g^{n}_{i})^2}{\mathrm{max}(g^{m}_{i} - g^{n}_{i})^2}
\Bigg]^{1/2},
\end{equation}
where $g^{m}_{i}$, $i=1,2,3,4$, are four Gini coefficients $g_{BC}^{m}$, $g_{CC}^{m}$, $g_{SC}^{m}$, $g_{IC}^{m}$, respectively, for the city $m$. We divided each distance by its maximum value to normalize the effect of each centrality because it has different scales.

We employed the complete-linkage clustering algorithm, which is a kind of hierarchical agglomerative clustering method, to construct the dendrogram. Initially, all of the nodes are in one cluster. The distance between two clusters is defined as the furthermost distance of the distances between two nodes from each cluster. At each step, the two clusters separated by the shortest distance are combined so that the cluster is formed. Next, we repeat this procedure until the clustering process results in a final state where all nodes are in the same single cluster. The dendrogram is obtained from the clustering procedure. Most relevant cities are placed in the neighborhood on the dendrogram.

\begin{figure*}[!t]
\includegraphics[width=0.95\textwidth]{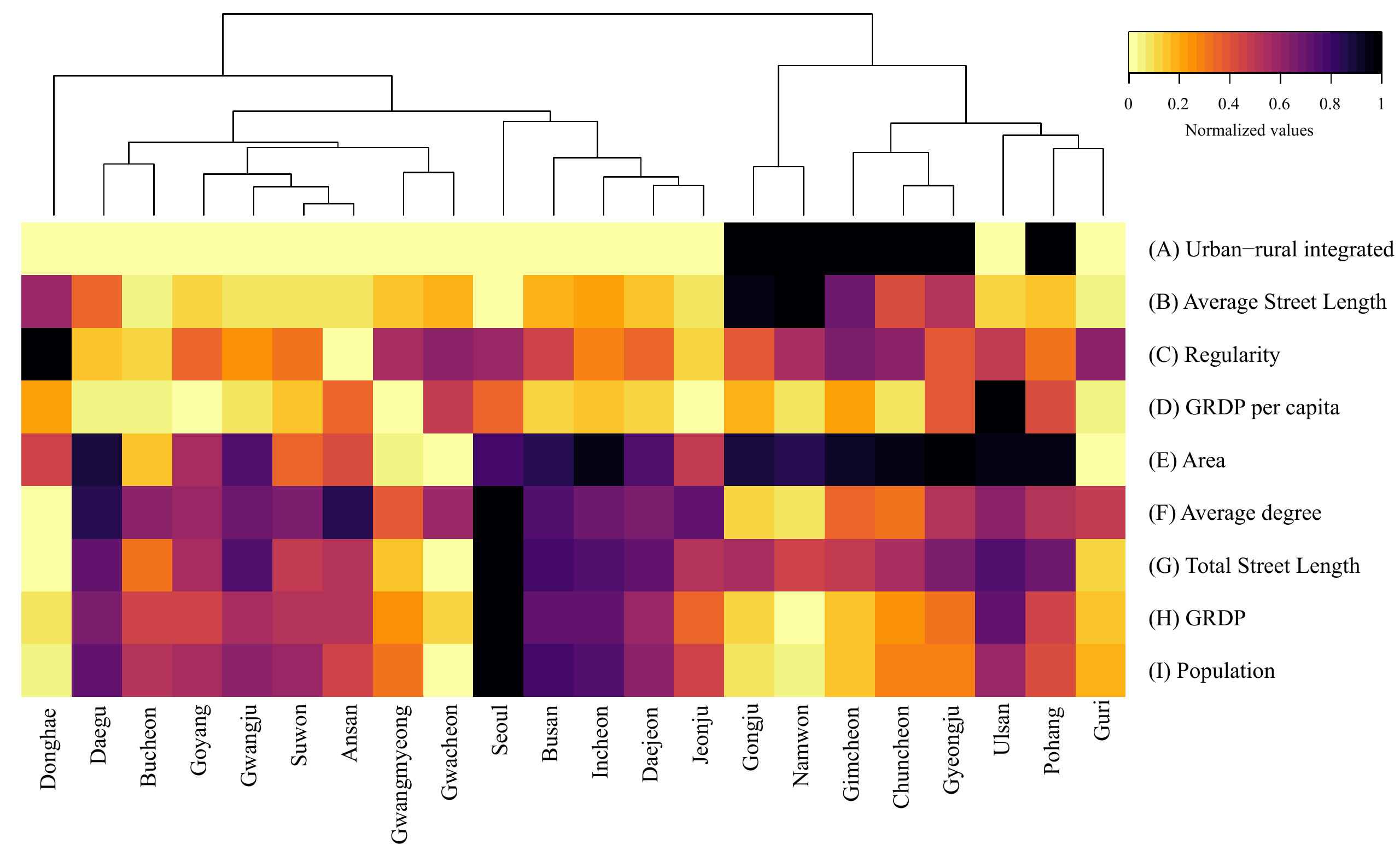}\centering
\caption{(Color online) The dendrogram and heat map for the 22 cities. The dendrogram is obtained based on the distance matrix in Eq. (7) of Gini coefficients calculated from the mixture of four kinds of centrality. The horizontal axis represents the cities that are hierarchically clustered and the vertical axis shows the meta-information of the cities. The quantities of the meta-information are normalized from zero to one for an easier comparison. The four kinds of the meta-information, (E), (G), (H), and (I), are normalized with the logarithmic scale because of the large range of their values. The cities in the same cluster at some levels have similar color patterns of the meta-information with a small exception.}
\label{fig:heat}
\end{figure*}

The classification result is shown in Fig. \ref{fig:heat}, and we also append the meta-information of the cities to clarify the relationship between the topology and meta-information. The horizontal axis represents the cities that are clustered by the above mentioned algorithm, and the vertical axis, from (A) to (I), shows the meta-information. There are two large distinct groups; one group comprises cities Donghae to Jeonju, and another group comprises cities Gongju to Guri. Each group has a similar color pattern, which means that the cities within the same group share similar external socio-economic characteristics. This kinds of logic is also valid in the several subgroups; the group comprising cities Goyang to Ansan, and the other group composed of cities Gwangmyeong and Gawcheon, for example.

Considering the classification microscopically, the cities are grouped into more specific traits. Because of the geographical factor, Gimcheon, Chuncheon, and Gyeongju are in the same group; these cities are provincial cities that has the large area. Seoul to Daejeon, which are metropolitan cities, are in the same group with exception of Jeonju. Gwacheon and Gwangmyeong, which are the small satellite cities of Seoul, are in the same group. Those kinds of meta-information of cities are reflected in the topology of street networks, and vice versa. The cities that have similar meta-information are located in the same subgroup of the dendrogram, with the implication being that there is a close connection between the topology and meta-information of the cities. Because the city is the dynamical system that is changed internally and externally by the time, the close relation between the street network and meta-information implies that they are simultaneously function as both the cause and effect in the development process of the city.

\section{Regularity in the Planned Cities}
The origin of the street structure is generally divided based on two cases. The first case is the bottom-up self-organized street structure; in this case, most of the streets are constructed without planning \cite{batty2007cities}. The second case is the top-down planned street structure, where the streets in the city are designed carefully to achieve a high efficiency flow of traffic, which reduces the serious congestion. As expected, many cities are combinations of those two organizations, but the degree of planning varies for each city.

The basic idea is that the cities that went through more planning policies have more regular street patterns. Regular patterns comprise a large proportion of four-way intersections, also called X-junctions, than three-way intersections, also called T-junctions. The X ratio and T ratio express the ratio of X-junctions and T-junctions, respectively \cite{marshall2004streets}. Here, we improve that ratios to reduce the influence of other intersections with a degree larger than four or smaller than two. We define the regularity $R$ of the street network as
\begin{equation}
R=\frac{P(k=4)-P(k=3)}{P(k=4)+P(k=3)},
\end{equation}
where $k$ refers to the degree of the nodes. $P(k=3)$ and $P(k=4)$ are the fractions of the nodes where the degree equals to three and four, respectively. The regularity is normalized by the sum of the fraction of nodes with degree three and four. Because there are more nodes where the degree equals three than the degree equals four, the values of regularity are generally negative.

\begin{figure}[!t]
\includegraphics[width=0.8\textwidth]{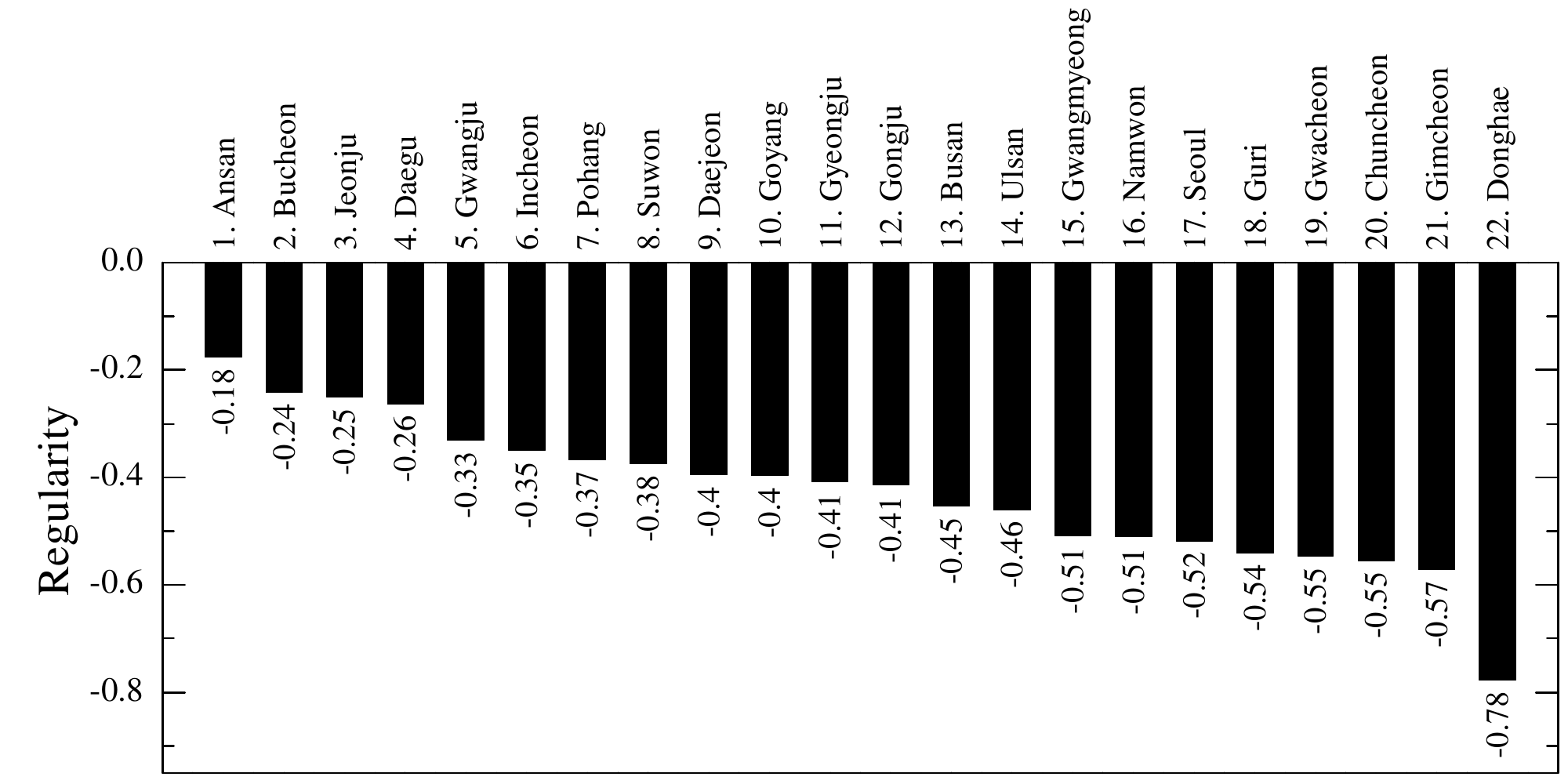}\centering
\caption{Most of the planned city are composed of the regular structure such as the grid-iron lattice in which nodes where degree equals four is relatively abundant. Regularity in Eq. (8) represents the proportion of the regular street structure in the city. When regularity is large, it means more regular pattern.}
\label{fig:reg}
\end{figure}

Fig. \ref{fig:reg} represents the values of the regularity for all cities which ranges from about -0.2 to -0.8. The larger the value of regularity, the closer it is to that of a regular planned city. For example, of the ten cities that have large values of regularity, seven cities contain districts built by top-down city planning \cite{newtown}, with the exception of three cities: Jeonju, Daegu, and Gwangju. In that sense, the street patterns in terms of the regularity are the outcomes of history of the top-down road planning. The regularity, which is a kind of topology, is related to the administrative policy, which is a kind of meta-information, in the planned cities.

\section{Scaling Law in Street System}
Cities have two quantities, the first of which is the social quantity, which is intangible but essential for economic activity that takes place within the cities. The second quantity is the infrastructure, which plays an indispensable role for the physical activity of the cities. If we denote the quantities of the city as $Y$, then $Y=Y_{0}N^{\beta}$, where $Y_{0}$ is a constant and $N$ is the population. Social quantities such as wages or new inventions follow the superlinear scaling \cite{bettencourt2007invention, bettencourt2007growth, bettencourt2010urban, bettencourt2013origins}, which means that the exponent $\beta$ is larger than 1, $\beta>1$. On the other hand, urban infrastructure such as roads or cables follow the sublinear scaling \cite{bettencourt2007invention, bettencourt2007growth, bettencourt2010urban, bettencourt2013origins}, which means that the exponent $\beta$ is smaller than 1, $\beta<1$.

\begin{figure*}[!t]\centering
\begin{subfigure}[t]{0.3\textwidth}\centering\caption{}
\includegraphics[width=\textwidth]{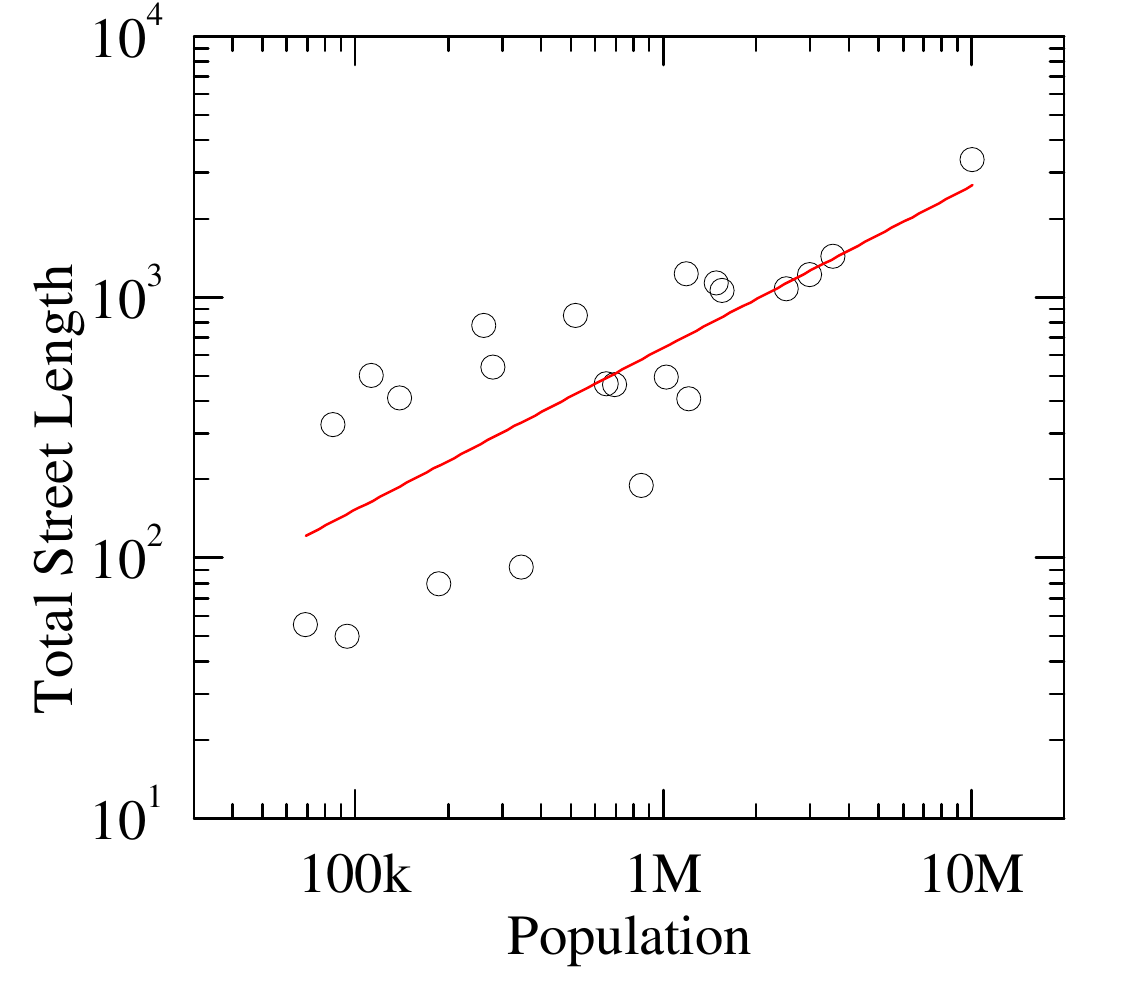}\label{scaling1}
\end{subfigure}\quad
\begin{subfigure}[t]{0.3\textwidth}\centering\caption{}
\includegraphics[width=\textwidth]{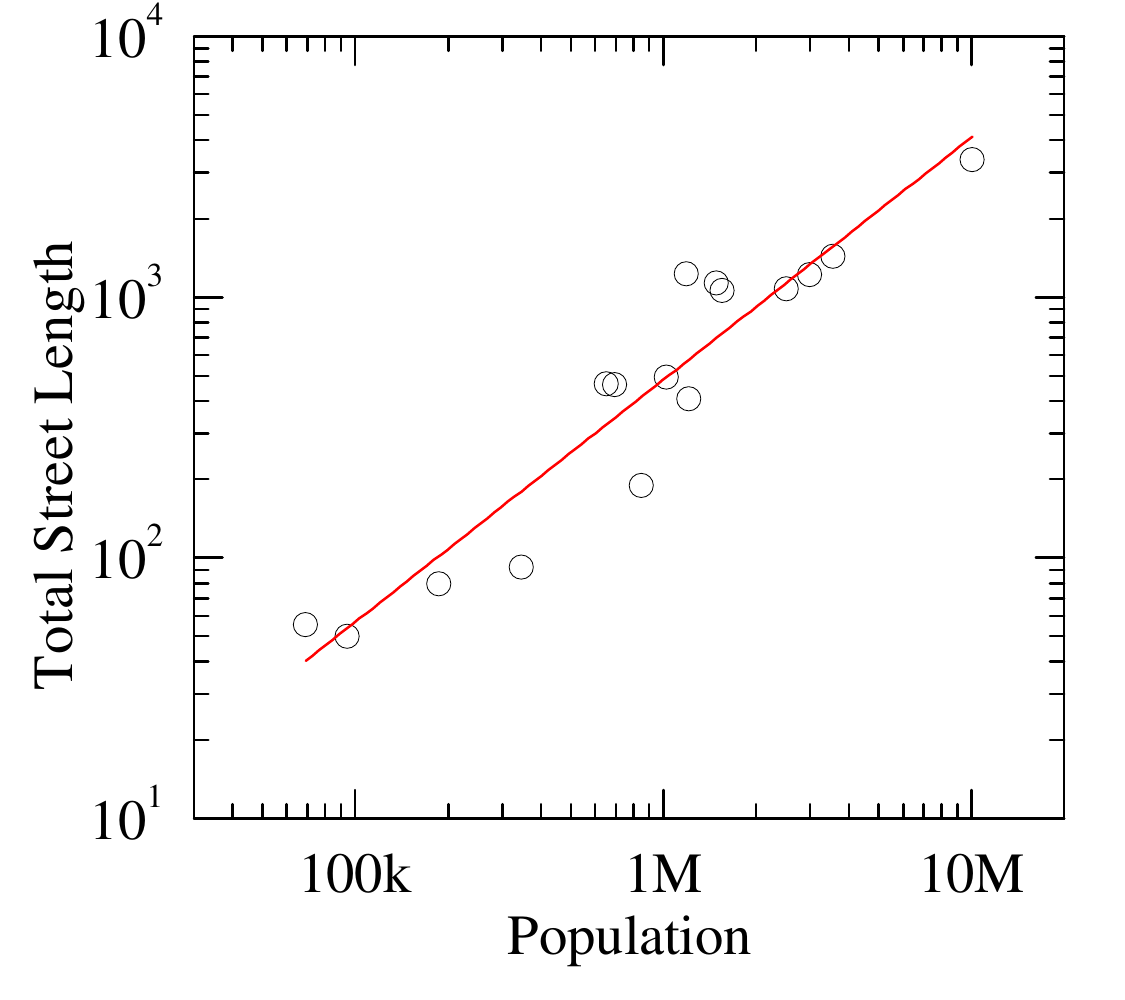}\label{scaling2}
\end{subfigure}\quad
\begin{subfigure}[t]{0.3\textwidth}\centering\caption{}
\includegraphics[width=\textwidth]{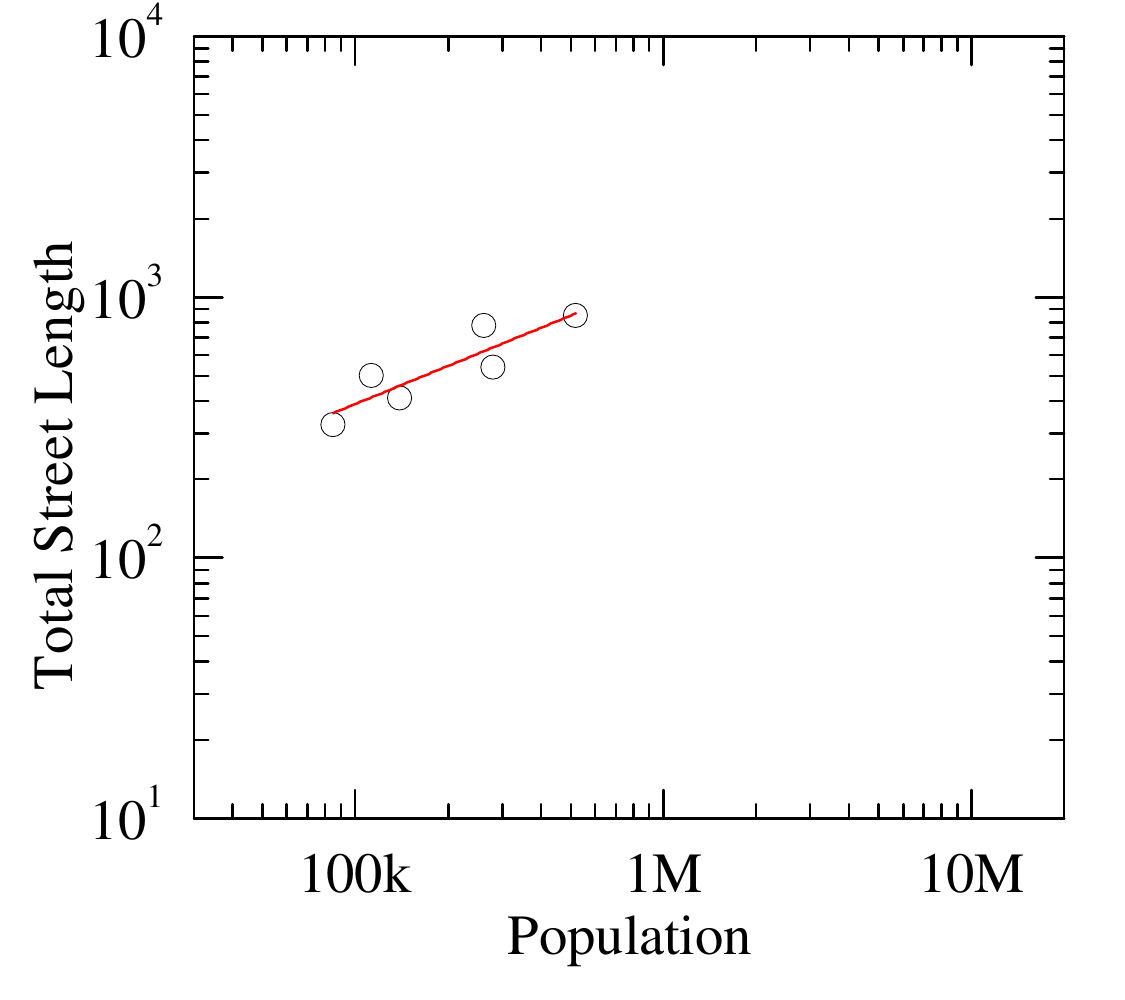}\label{scaling3}
\end{subfigure}
\caption{The scaling relation is $L \, \sim \, N^{\beta}$, where $L$ is the total street segment length and $N$ is the population of the city. The exponent $\beta$ is generally less than 1 for infrastructure. (a) Scaling law of the total street segment length in $\mathrm{km}$ for all 22 cities. The exponent $\beta$ is 0.599 ($adj. R^{2} = 0.4673$). (b) Scaling law for 16 cities that are not urban-rural integrated. The exponent $\beta$ is 0.930 ($adj. R^{2} = 0.8902$). (c) Scaling law for 6 urban-rural integrated cities. The exponent $\beta$ is 0.485 ($adj. R^{2} = 0.7367$), which is smaller than that of the usual cities in (b).}
\label{fig:scaling}
\end{figure*}

In this section, we obtain that the road system of Korea follows the sublinear scaling with some notable characteristics, as shown in Fig. \ref{fig:scaling} (a). In the case where we consider all of the cities, the relationship between the population and the total street segment length follows the sublinear scaling with exponent $\beta=0.599$.

We divide the cities into two categories. The first group comprises usual cities where whole sections of the cities are closely connected by built-up areas, and the second category comprises urban-rural integrated cities, which is a combination of urban areas and rural areas that are apart from each other in the early stage. The purpose of the urban-rural agglomeration is primarily for administrative convenience, but in doing so, the two parts of the city are closely connected economically. The urban-rural integrated cities exhibit a typical core-periphery structure. The density of the roads and intersections are dense in central areas, but sparse in fringe areas. Here, we do not rigorously classify the two structures as core-periphery and non-core-periphery, but most of the urban-rural integrated cities have core-periphery characteristics.

Fig. \ref{fig:scaling} (b) shows the sublinear scaling relation for cities that are not urban-rural integrated cities, and the exponent is $\beta=0.930$, which is somewhat large for infrastructure but which is the sublinear scaling. Fig. \ref{fig:scaling} (c) represents the sublinear scaling relation for the only urban-rural integrated cities. The exponent is $\beta=0.485$, which is a very small value. The difference in the exponent implies that urban-rural integrated cities are less influenced by the change of the population because they already have a sufficient number of streets in the periphery rural region. As a result, the citizens living in urban-rural integrated cities tend to receive relatively fewer benefits from the population bulge. This is an example for which administrative policy affects the landscape of the street structure and the degree of importance of the streets as an essential form of infrastructure of the cities.

\section{Conclusion}
Street structures, which are the central infrastructure of the city, coevolve along with the social and economic quantities of the city. While the economy of a city expands so that its gross regional domestic product increases, new streets are constructed using the increased funding. Once the streets are expanded and densified, then the efficiency of transportation of the goods and people increases. Consequently, their economic state again expands upon using the improved infrastructure. Therefore, it is important to understand the underlying mechanisms of this cycle as well as the interdependent relationship between the street structure and meta-information of the city.

In this study, we analyzed the street structure using the four types of centrality: betweenness, closeness, straightness, and information centrality. Of the four centralities, the value of the betweenness centrality has the largest heterogeneity. Heterogeneity is key to the classification of cities. Based on the Gini coefficients of those centralities, we grouped the 22 cities in Korea by their topological relevance. The normalized values of meta-information, such as the population or gross regional domestic product, are attached in the dendrogram. We found that the classification according to the topology of the street network is well matched with the heat map constructed by the meta-information of the city, which implies that there are close associations between them.

Top-down city planning is another topic, and it considers the construction of street structures such that there is a reduced occurrence of serious traffic jams and increased traffic efficiency, which are the major issues pertaining to road planning. We define the regularity to measure how well they are planned from a street perspective. Finally, the cities are lined up consistently with the planned city in reality.

The scaling relation between the population and the street segment length is sublinear. If we divide the cities into two categories, i.e., usual cities and urban-rural integrated cities, then we obtain different exponents, which indicate that the street structures of urban-rural integrated cities are less influenced by the change of the population. Therefore, the administrative policy of the city is another factor that has to be considered to gain a deeper understanding of evolution of the city.

There remain some open questions with respect to the centrality analysis. The first one is the fundamental problem of the centrality itself. For example, the betweenness centrality is not an exact proxy for the traffic volume. When calculating the betweenness centrality, we usually treat all of the nodes as having the same weight, but each node gives different traffic volume that is proportional to the number of car registrations at that spot or the floating population. In addition, in reality, drivers do not usually stick to only the shortest path; instead, they utilize faster and more convenient roads such as highways, and have their own standards for choosing the path. Alternately, navigation devices can disturb paths. Furthermore, traffic lights located at the different intersections should be considered when calculating the shortest path. These factors are intrinsic limitations of the betweenness centrality. If we employ other forms of improved centrality, such as the random walk betweenness centrality \cite{newman2005measure}, which considers a larger number of factors to determine the optimal path, then the result may be more accurate.

Another point is that there is no universal definition of a city. In this paper, we treat a city from an administrative perspective. However, cities may be defined in various ways such as using the population density threshold or commuting threshold \cite{arcaute2015constructing}. From a traffic perspective, it is more appropriate to define a city as a commuting zone. Although there would be commuters who cross the city boundaries every day, we assume that their numbers are negligible within the administrative boundary of the city. In our case, 13 cities are in the local region in Korea, and in those cities, there are no significant connections with the neighboring towns compared to the internal communication; thus, a large proportion of their meta-information, such as the gross regional domestic product, are limited to within the boundary of the cities. On the other hand, nine cities, including Seoul, are within the capital region, which means that the cities interact actively with each other. As a result, the scale of their meta-information is underestimated or overestimated by the overlapping, and this could therefore distort the relationship between the meta-information and the street structure. In future, we will conduct further studies that consider the effective boundaries of cities, with the aim of enhancing our understanding of the evolution of cities.

\section*{Acknowledgments}
This work was supported by Basic Science Research Program through the National Research Foundation of Korea (NRF) funded by the Ministry of Education (NRF-2016R1D1A1B03932590) and Basic Science Research Program through the National Research Foundation of Korea (NRF) funded by the Ministry of Education (NRF-2016K2A9A2A08003695).

\section*{References}


\bibliographystyle{elsarticle-num} 

\end{document}